\documentclass[
                aip, jcp,
                preprint,
                superscriptaddress,
                amsmath,
                amssymb,
                floatfix
                ]{revtex4-1}
 \usepackage{amsfonts}
 \usepackage{amsmath}
 \usepackage{graphicx}
 \usepackage{subfigure}
 \usepackage[colorlinks=true,
             linkcolor=blue,
             citecolor=blue,
             filecolor=blue,
             urlcolor=blue]{hyperref}

\renewcommand{\vec}[1]{\mathbf{#1}}

\begin{document}

\title{Sub-picosecond energy transfer from a highly intense THz pulse to water:
       a computational study based on the TIP4P/2005 model.}

\author{Pankaj Kr. Mishra}
\email[]{pankaj.kumar.mishra@desy.de}
\affiliation{%
    Center for Free-Electron Laser Science, DESY,
    Notkestra\ss e 85, D-22607 Hamburg, Germany}
\affiliation{%
    The Hamburg Centre for Ultrafast Imaging,
    Luruper Chaussee 149, D-22761 Hamburg, Germany}
\affiliation{%
    Department of Physics, University of Hamburg,
    Jungiusstra{\ss}e 9, D-20355 Hamburg, Germany}
\author{Oriol Vendrell}
\email[]{oriol.vendrell@cfel.de}
\affiliation{%
    Center for Free-Electron Laser Science, DESY,
    Notkestra\ss e 85, D-22607 Hamburg, Germany}
\affiliation{%
    The Hamburg Centre for Ultrafast Imaging,
    Luruper Chaussee 149, D-22761 Hamburg, Germany}
\author{Robin Santra}
\affiliation{%
    Center for Free-Electron Laser Science, DESY,
    Notkestra\ss e 85, D-22607 Hamburg, Germany}
\affiliation{%
    The Hamburg Centre for Ultrafast Imaging,
    Luruper Chaussee 149, D-22761 Hamburg, Germany}
\affiliation{%
    Department of Physics, University of Hamburg,
    Jungiusstra{\ss}e 9, D-20355 Hamburg, Germany}

\begin{abstract}
The dynamics of ultrafast energy transfer to water clusters and to bulk
water by a highly intense, sub-cycle
THz pulse of duration $\approx$~150~fs is investigated in the context of
force-field molecular dynamics simulations.
We focus our attention on the mechanisms by which rotational and translational
degrees of freedom of the water monomers gain energy from these sub-cycle
pulses with an
electric field amplitude of up to about 0.6~V/{\AA}. It has been recently shown
that pulses with these characteristics can be generated in the laboratory [PRL
112, 213901 (2014)].
Through their permanent dipole moment, water molecules are acted upon by the electric
field and forced off their preferred hydrogen-bond network conformation.
This immediately sets them in motion with respect to one another as energy
quickly transfers to their relative center of mass displacements.
We find that, in the bulk, the operation of these mechanisms is strongly
dependent on the initial temperature and density of the system.
In low density systems, the equilibration between rotational and translational
modes is slow due to the lack of collisions between monomers.
As the initial density of the system approaches 1~g/cm$^3$, equilibration
between rotational and translational modes after the pulse becomes more
efficient.
In turn, low temperatures hinder the direct energy transfer from the pulse to
rotational motion owing to the resulting stiffness of the hydrogen bond network.
For small clusters of just a few water molecules we find that fragmentation
due to the interaction with the pulse is faster than equilibration between rotations
and translations, meaning that the latter remain colder than the former after
the pulse.
In contrast, clusters with more than a few tens of water molecules already
display energy gain dynamics similar to water in condensed phases owing to
inertial confinement of the internal water molecules by the outer shells. In
these cases, a complete equilibration becomes possible.
\end{abstract}

\maketitle

\section{Introduction}

Water in its liquid form is the universal solvent in which most chemical
processes in a biological environment take place. For this reason, its
properties under variations of thermodynamic parameters such as
temperature~\cite{Smith14171} and pressure~\cite{Schwegler2429}, under external
perturbations such as electric fields~\cite{Jung331,saitta207801} and
ultrasound waves~\cite{Dahl677}, or under
confinement~\cite{Zangi1694,Evans7138,Hirunsit1709,marti21504,Gordillo341,
Rasaiah713} have been the subject of much attention.
%
%
Isolated water molecules have a permanent dipole moment of
1.85~D~\cite{Clough2254}, which varies with cluster size and converges to 2.85~D
for liquid water~\cite{Silvestrelli3308}.
In the presence of an electric field, the permanent dipole of the water monomers
experience a torque, which tries to orient the water dipole
along the polarization direction of the field and perturbs the equilibrium
arrangement formed by the hydrogen bond (H-bond) network~\cite{Vegiri8786}.
For example, a DC or slowly varying electric field with an amplitude in the range of $0.5\times
10^{7}$-$10\times 10^{7}$~V/cm can induce structural transformations in water
clusters and modify their equilibrium geometry from ring-like to chain-like
structures ~\cite{choi94308,Vegiri4175,Rai34310,buck2578,shevkunov27,choi94308}.
These changes are linked to a reduction of the average number of H-bonds in the
system.
Strong electric fields polarize the electronic cloud and lead to modifications
of the dipole moment of the water monomers. These changes are dependent on
cluster size and field strength~\cite{gregory814}.
Other static properties of bulk water like its diffusion
constant~\cite{Vegiri4175,Vegiri8786} or its intramolecular vibrational
frequencies~\cite{Rai34310} can also be altered by the structural modifications
caused by the presence of electric fields.
Electric fields can also have an effect on the mechanism
of biological processes occurring in liquid water.
Examples include the modification of rate constant of folding and unfolding
processes of proteins and peptides~\cite{Lockhart947,Freedman5137,Fox5277}
and of other types of chemical reactions~\cite{Franzen5135,Gopher311}.

Laser sources have also been applied to induce
structural~\cite{bakker_reorientational_2000,link_ultrafast_2009}
and dynamical~\cite{shimano_intense_2012,rinne_nanoscale_2012}
modifications in liquid water.
%
Femtosecond laser pulses can be tuned to the excitation of internal degrees of
freedom of the water
monomers, e.g. vibrational modes~\cite{imoto_ultrafast_2015,woutersen507}.
Specifically, infrared (IR) lasers can directly excite the intramolecular O-H
stretching modes, which strongly absorb at wavenumbers $1/\lambda\approx
3500$~cm$^{-1}$. After the pulse, the energy deposited by
the laser pulse relaxes via vibrational energy
redistribution to other intramolecular and intermolecular vibrational
modes~\cite{Woutersen1997,woutersen658,hasted622,huse389}.
%
%
Alternatively, sub-cycle and highly intense THz pulses can also transfer a large amount of energy
to liquid water, which can be the equivalent of several
hundreds of degrees Kelvin~\cite{mishra51,MishraJCPB}.
The mechanism by which the sub-cycle THz pulse interacts with the medium is
off resonance with intramolecular degrees of freedom and with librational modes
of bulk water~\cite{MishraJCPB} and more closely
related to the interaction mechanism of water molecules with DC fields.
%

In this work, we investigate the interaction of intense, sub-cycle
THz pulses with
water clusters of various sizes and with bulk water at a range of initial temperatures and
densities. The pulses considered are
characterized by peak electric field amplitudes of up to 0.6~V/{\AA}. Pulses of similar peak field amplitude in the frequency range of
1-10~THz have recently been demonstrated using non-linear organic
crystals~\cite{vic14:213901}.
Our simulations are based on the TIP4P/2005 rigid water
model~\cite{aba05:234505}.
Compared to preceding studies of THz-pumped liquid water based on {\em ab
initio} molecular dynamics (AIMD) simulations~\cite{mishra51,MishraJCPB}, which
were performed on much smaller simulation boxes and for very specific initial
conditions, the use of a parametrized force field allows
for a systematic investigation of the transition from cluster to bulk behavior
and of a wide range of initial temperatures and densities.
We first concentrate on the interaction of the THz pulse with one isolated water
molecule and with clusters of a few water molecules. Based on these simulations
we discuss the basic mechanism by which the THz pulse initially transfers energy
to the system and how this energy is transferred among the hindered rotational
and translational modes of nearby interacting water molecules.
Subsequently, we discuss the interaction of bulk water at various initial
temperatures and densities under periodic boundary conditions at
constant volume and particle number. This means, essentially, that some form of
external confinement of the water molecules in the interaction volume with the
THz pulse is being assumed.
This can correspond to a macroscopic confinement of the whole interaction region by
hard walls that keep the system from expanding as it heats up,
or it can correspond to inertial confinement of the region of a bulk system
crossed by the pump pulse.
By simulating clusters with up to several thousands of water molecules, we
establish that indeed the system is inertially confined as it heats up. Since the
energy intake proceeds on a sub-picosecond time scale, the inner regions of the
cluster remain essentially at their initial density throughout the confinement exerted
by the outer shells, which need a longer time to start moving away from the
cluster.
For periodic systems at low density and high temperature, the behavior of
isolated water molecules is recovered, whereas at a density of 1~g/cm$^3$ the
amount of energy gained by the system is found to be strongly dependent on the
initial temperature, which determines the average strength of the hydrogen
bonds.

The structure of the paper is as follows. Section~\ref{sec:comp} describes the
computational methods. Section~\ref{ssec:clusters} discusses the interaction of
the THz pulse with water clusters and the transition to bulk behavior.
Section~\ref{ssec:bulk} describes the response of bulk water at different
initial temperatures and densities to the THz pulse, and Section~\ref{sec:concl}
concludes and provides some outlook.

\section{Computational details}
\label{sec:comp}


Simulations of the water clusters and of bulk water under periodic boundary
conditions were performed with the Large-scale Atomic/Molecular Massively
Parallel Simulator (LAMMPS)~\cite{pli95:1} using the rigid TIP4P/2005 water
force field model~\cite{aba05:234505}.
For the water cluster simulations, 1000 initial atomic configurations and
velocities were generated from a long molecular dynamics (MD) trajectory
performed under canonical (NVT) conditions and equilibrated at 200~K. Different
sets of initial conditions were sampled with a 10~ps time interval to avoid
artificial correlations between each sampled phase space point.
The Nose-Hoover thermostat~\cite{evans_nosehoover_1985,nose_extension_1986} was
used to maintain the temperature for the phase space sampling NVT trajectory.
Each set of 1000 initial conditions was then propagated microcanonically for
1.5~ps in the presence of the THz pulse.
For all trajectory propagations the velocity-Verlet algorithm was used with a
time step of 1~fs.
A cut-off radius of 15~{\AA} was used for the calculation of the short-range
Lennard-Jones interactions.
Water clusters of sizes 1, 2, 4, 8, 32 and 64 monomers were
initially considered.
In order to study the effect of inertial confinement, a bigger spherical cluster with 8843
water molecules and radius $\approx$~50~{\AA} was equilibrated at
300~K and subsequently
exposed to the THz pulse.
All water cluster simulations were performed with non-periodic boundaries in all
directions.

Bulk water simulations at different initial temperatures and densities were
based on thermally equilibrated cubic boxes containing 2048 water molecules.
By selecting the box size, a range of initial densities
of 1.0~g/cm$^3$, 0.296~g/cm$^3$, 0.064~g/cm$^3$ and 0.019~g/cm$^3$ were
considered. Each box was then equilibrated at different temperatures of
200~K, 300~K, 400~K, 500~K and 600~K.
Each set of initial density and temperature conditions was subsequently propagated
microcanonically for 1.5~ps in the presence of the THz pulse.

The same half cycle THz pulse profile as in previous
investigations was considered here~\cite{mishra51,MishraJCPB}.
The THz pump pulse is given by
\begin{equation}
 \label{pump-pulse}
 \mathbf{E}(t) = \epsilon(t)\mathbf{u}_{z}\cos(\omega_{c}t+\phi),
\end{equation}
where $\epsilon(t) = A\exp\{-(t-t_{0})^{2}/2\sigma^{2}\}$ is
a Gaussian envelope with $\sigma = 84.93$~fs.
This corresponds to a full width at half maximum (fwhm)
of $\epsilon^{2}(t)$ of 141 fs.
The maximum electric field amplitude $A = 0.61$~V/{\AA} corresponds to a
power per unit area of $5\times10^{12}$~W/cm$^{2}$.
The mean photon frequency considered is
$\omega_c= 2\pi\times 3$~THz (100~cm$^{-1}$) which results in a pulse
between a half and a full cycle long.
$\mathbf{u}_{z}$ is the
polarization direction of the electric field and $\phi$ is the carrier to
envelope phase which is set to $\pi/2$.
By convention, the THz pulse envelope is centered at $t_{0}= 0$ in all
simulations and microcanonical trajectories start at $t = -250$~fs.

The analysis of the energy gain by water clusters and bulk water is
performed on the basis of decomposing the total kinetic energy ($E_K^{(m)}$) of each
rigid water monomer $m$ in terms of its rotational energy ($E_R^{(m)}$) and translational
energy ($E_T^{(m)}$) components.
This is simply achieved in Cartesian coordinates by,
\begin{align}
    \label{eq:traE}
    E_{T}^{(m)} & = \frac{ |\sum_{a_m} M_{a_m} \vec{V}_{a_m}|^2  }
                     {2 \sum_{a_m} M_{a_m}}
\end{align}
and
\begin{align}
    \label{eq:rotE}
    \nonumber
    E_{R}^{(m)} & = \sum_{a_m} \frac{\vec{j}_{a_m}^2}{2 M_{a_m} |\vec{x}_{a_m}|^2}\\
    \vec{j}_{a_m} & = M_{a_m}\, (\vec{x}_{a_m} \times \vec{v}_{a_m}),
\end{align}
where $\vec{x}_{a_m} \equiv \vec{X}_{a_m} - \vec{X}_m$ and
$\vec{v}_{a_m} \equiv \vec{V}_{a_m} - \vec{V}_m$ define the position and
velocity of atom $a$ in monomer $m$ relative to the center of mass position
$\vec{X}_m$ and velocity $\vec{V}_m$ of the monomer, respectively.
In general, at this point the kinetic energy per monomer related to vibrational
motion of the remaining $3N-6$ intramolecular coordinates $E_V^{(m)} = E_K^{(m)}
- E_T^{(m)} - E_R^{(m)}$ can be
obtained. In the present case, owing to the rigid nature of the water model
used,
$E_V^{(m)}=0$. In particular, the intramolecular degrees of freedom
were kept fixed in LAMMPS by the
SHAKE algorithm~\cite{ryc77:327}.

\section{Results and Discussion}
\label{sec:results}

\subsection{Interaction of the THz pulse with water clusters}
\label{ssec:clusters}

\subsubsection{Water monomer and dimer}

\begin{figure}[h!]
      \includegraphics[width=9cm]{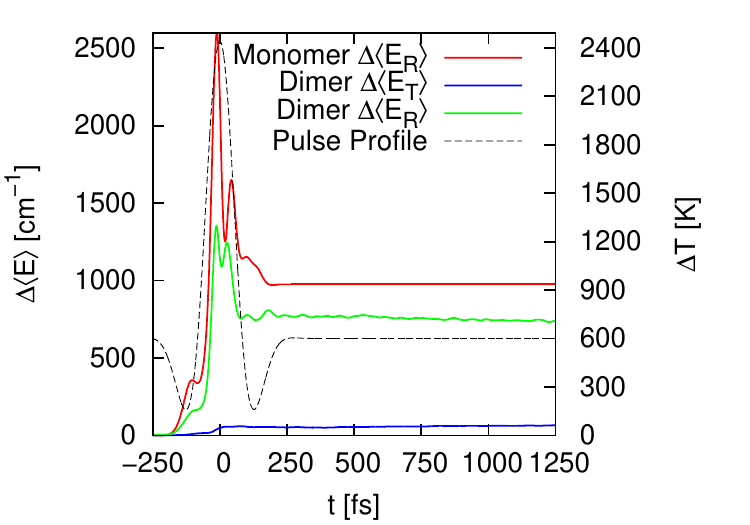}
      \caption{Translational ($\Delta\langle E_T\rangle$)
          and rotational ($\Delta\langle E_R\rangle$) energy
          increase per water monomer for the monomer and dimer cases.
          On the right axis the kinetic temperature $T=\frac{2 E}{3 k}$ with
          $k$ the Boltzmann factor is shown.
          The THz pulse profile appears as a dotted line.
          }%
      \label{fig:water-monomer-energy}
\end{figure}
The effect of the THz pulse on a single water monomer and on the water dimer is
analyzed first. The ensemble averaged translational $\langle E_T\rangle$ and
rotational $\langle E_R\rangle$ energies per monomer were obtained by
averaging over 1000 initial configurations and velocities. In the case of the
monomer and due to the intramolecular constraints, the sampling of
initial conditions
generates only random orientations and angular velocities of the water
molecule, the latter being compatible with a target temperature of 200~K. In the
dimer case, overall rotations of the system as well as intermolecular coordinates
are sampled according to the target temperature of 200~K.
The electric field is polarized along the z-axis in all
simulations.
Due to the electroneutrality of the isolated water monomer, only the rotational
degrees of freedom gain energy in the presence of the THz pulse, as illustrated
in Fig.~\ref{fig:water-monomer-energy}.
The time structure of the pulse when it changes its polarization direction can
still be seen in $\Delta\langle E_R\rangle$ in spite of the rotational averaging over
initial conditions.
As expected, no further change in total energy occurs after $t=250$~fs, when the
pulse is over.

In the dimer case, both $\Delta\langle E_R\rangle$ and $\Delta\langle E_T\rangle$ increase
due to the THz pulse. However, the increase of translational energy of the
monomers is, in this, case much smaller than that of rotational energy.
On average, the dimer fragments in its two water molecule constituents
very early during or shortly after the THz pulse, which closes the channel for
energy transfer between rotational and translational motion by interactions
between the monomers. Therefore, the relative translational mode between both
centers of mass remains much colder than the individual rotational modes of both
monomers, and equipartition cannot be reached.
To understand the origin of the translational energy increase of the monomers we
can first inspect the form of the interaction term between matter and radiation.
In the non-polarizable water model considered here the
interaction of the electric field with the system is simply the sum of the interactions with the
individual dipole moments of the monomers
\begin{align}
    \label{eq:energy-1112}
    E^{w-EF} = - \sum_{m} \sum_{a_m}
                 q_{a_m} ( \vec{r}_{a_m} \cdot \vec{E} )
             = - \sum_{m}\boldsymbol{\mu}_m\cdot \vec{E},
\end{align}
where
$\vec{r}_{a_m}$ is the position of the $a$-th point charge of the $m$-th
monomer, $\boldsymbol{\mu}_m$ is the permanent dipole moment of monomer $m$ and
$\vec{E}$ is the external electric field.
This term has no {\em direct} effect on the translational motion of the centers
of mass and only depends on the orientation of each individual dipole in the
electric field.
The total potential energy ($E_{P}$) of the system in the presence of the
electric field consists of the sum of pair interactions between monomers and the interaction
of each monomer with the field
\begin{align}
    \label{eq:energy}
    E_P = E_{\text{pair}}^{w-w} + E^{w-EF}.
\end{align}
$E_{\text{pair}}^{w-w}$ includes a Lennard-Jones term between each water
molecule, which is spherically symmetric, and Coulombic interactions between
each effective partial charge in each water molecule, which account for the
effective charge
distribution in the water model~\cite{aba05:234505}. This last part of
$E_{\text{pair}}^{w-w}$ is strongly directional and is responsible for the
adequate
description of the coordination and H-bonding between water molecules.
Hence, the rotation of the water monomers forced upon them by the electric field through
$E^{w-EF}$ results in unfavorable configurations due to the $E_{\text{pair}}^{w-w}$
term, which leads to translational motion of the centers of mass of the monomers
and to energy transfer between monomer rotations and translations.
Higher order multipolar interaction terms than dipole-dipole
are embedded in the $E_{\text{pair}}^{w-w}$ term and are crucial to describe the
properties of liquid water
~\cite{abascal15811,niu134501,Patey170,vega1361}.
However, for the sake of argument let us consider for a moment
a simplified model of permanent point dipoles.
%
The potential energy of a system of permanent point dipoles reads
\begin{align}
    \label{eq:10}
     E_{\text{pair}}^{d-d} = \sum_{i>j}-\frac{{\mu}_{i}{\mu}_{j}}
                    {r_{ij}^3}
                    (\cos\theta_{ij}-3\cos\theta_{i} \cos\theta_{j})
\end{align}
where $r_{ij}$ is the distance
between the centers of the two dipoles, $\theta_{ij}$ is the
relative angle between the dipoles and
$\theta_{i}$ and $\theta_{j}$ are the
angles formed by the two dipoles with respect to the line connecting
their centers.
Since the dipole-dipole interaction term couples the relative orientation and
relative
distance of each pair of dipoles, an external electric field can transfer energy to the center of
mass translational motion of the individual monomers via the indirect mechanism
described above.
In the case of aggregates of molecules with individual permanent dipoles, this
constitutes the zeroth order mechanism for energy transfer from an
external field to the relative
translational motion of each constituent through their individual rotational
degrees of freedom. Polarization effects are not considered at this level
in such a mechanism.
It should be emphasized that the pulse frequency of 3~THz
is smaller by more than one order of magnitude than intramolecular
vibrational modes of separate water monomers and about 5 times smaller than hindered
librational modes of water monomers~\cite{heyden12068}. Hence, the interaction with the
THz pulse occurs in a non-resonant fashion as a response to the
instantaneous electric field amplitude.

\begin{figure}[t!]
    \begin{center}
            \includegraphics[width=8cm]{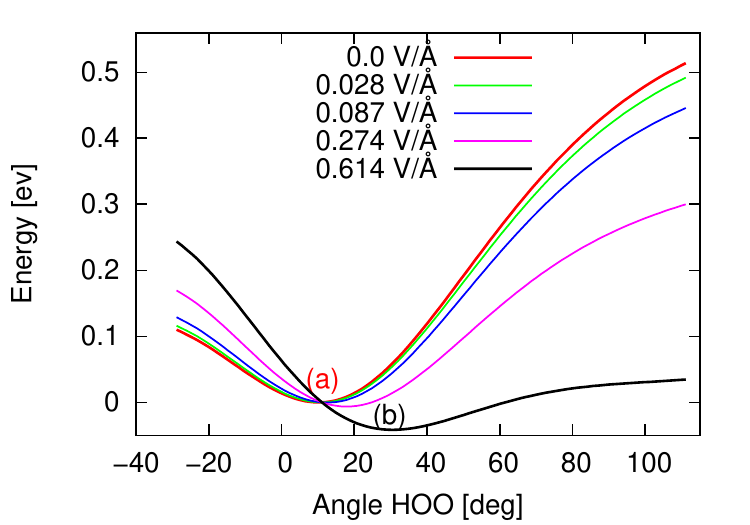}
    \end{center}
     \caption{Potential energy curve for
              HOO angle with different electric field amplitudes.
              The labels (a) and (b) correspond to the dimer configurations
              as shown in Fig.~\ref{fig:dimer-snapshot}.
      }%
      \label{fig:potential-energy-curve}
    \end{figure}

The effect of an electric field on the preferred relative orientation between the two water monomers
is clearly seen by inspecting the potential energy curve of the water
dimer along the $\angle$HOO relative angle in
Fig.~\ref{fig:potential-energy-curve}. The potential energy curves correspond to the
TIP4P/2005 water force-field used in the trajectory propagations.
    \begin{figure}[t!]
      \begin{center}
         \subfigure[]{%
            \label{fig:snap-equi-angle}
            \includegraphics[width=8cm]{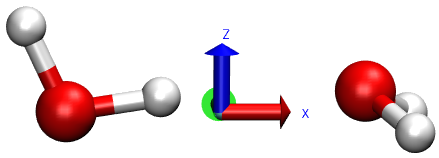}
         }\\%
         \subfigure[]{%
            \label{fig:snap-equi-angle-field}
            \includegraphics[width=8cm]{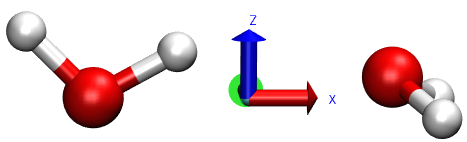}
         }%
     \end{center}
     \caption{Minimum energy configurations of the water dimer (a) without field and
         (b) with a field of 0.614 V/{\AA} as presented in  Fig.~\ref{fig:potential-energy-curve}.
      }%
    \label{fig:dimer-snapshot}
    \end{figure}
In the minimum energy configuration of the dimer, two water monomers
orient themselves to form a H-bond.
The applied static electric field is polarized along the x-axis, which is
the direction of the H-bond in the minimum energy configuration of the dimer
shown in Fig.~\ref{fig:snap-equi-angle}.
To construct the potential energy curves along the $\angle$HOO, the donor water molecule is
rotated in the XZ plane around the y-axis.
The potential energy curves were compared for static electric fields with
amplitudes 0, 0.028, 0.087, 0.274 and 0.614 V/{\AA}, which in the case of a
pulse would correspond to intensities of
0, 10$^{10}$, 10$^{11}$, 10$^{12}$ and 5$\times$10$^{12}$~W/cm$^{2}$.
The change in the shape of the potential is the result of the relative magnitude
of the
terms $E_{\text{pair}}^{w-w}$ and $E^{w-EF}$ in Eq.~\ref{eq:energy}.
By changing the field amplitude from 0.274~V/{\AA} to  0.614~V/{\AA},
the shape of the potential energy curve changes significantly in comparison to lower amplitudes.
Such high electric field is able to create a displaced and deep potential well
along the relative orientation between the water molecules.
At the high field amplitude, the electric field is strong enough to substantially alter the
H-bond and bring the water molecules out of their preferred field free
relative orientation.
The time delay of about 100~fs in the rise of the translational component
$\langle E_T\rangle$ compared to the start of the pulse and seen in
Fig.~\ref{fig:water-monomer-energy} is related to the fact that
the electric field has to reach a certain peak amplitude in order to disrupt the H-bond and
push the water molecules into configurations for which their relative distance
will start to change.
%

The H-bond strength of the water dimer is 23.8~KJ/mol ($\approx$~0.25~eV)~\cite{Urbic159}.
As a consequence, a certain threshold value $\epsilon_T$ for the
external electric field 
amplitude is required in order to bring water molecules out of their preferred
H-bond arrangements~\cite{choi94308,Rai34310}, as already seen in
Fig.~\ref{fig:potential-energy-curve}.
The THz pulse can then be divided into different parts, depending on whether the
electric field amplitude $\epsilon(t)$ is larger or smaller than the threshold value $\epsilon_T$
required to disrupt the H-bond structure.
From Fig.~\ref{fig:potential-energy-curve} one can estimate $\epsilon_T$ to be of the
order of $0.2$ to $0.3$~V/{\AA}.
Only when $\epsilon(t) > \epsilon_{T}$, which for the half cycle pulse
of Eq.~\ref{pump-pulse} is achieved only in the central oscillation,
the electric field can effectively bring molecules out of
their H-bond arrangements.
This is the reason why, already for the dimer case, $\Delta\langle E_R\rangle$
increases mostly during the central pulse oscillation and hardly changes
during the two side wings of smaller amplitude.

\subsubsection{Water tetramer and larger clusters}

    \begin{figure}[t!]
      \begin{center}
         \subfigure{%
            \label{fig:cluster-bulk-energy}
            \includegraphics[width=8cm]{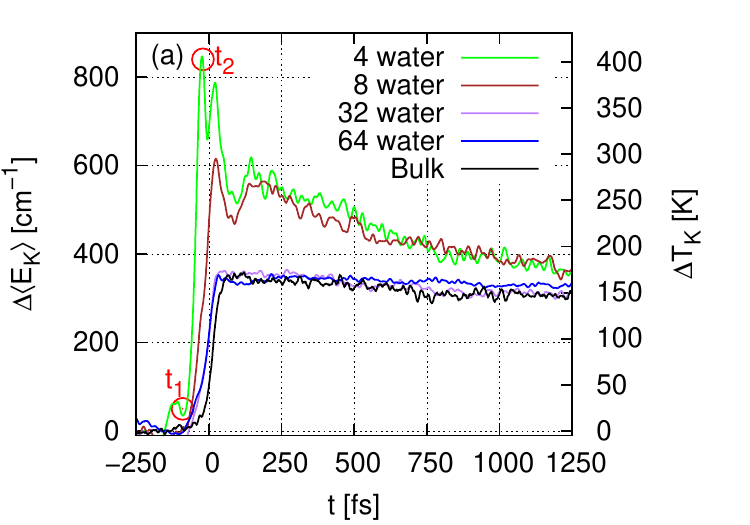}
         }\\%
         \subfigure{%
            \label{fig:cluster-bulk-trans-energy}
            \includegraphics[width=8cm]{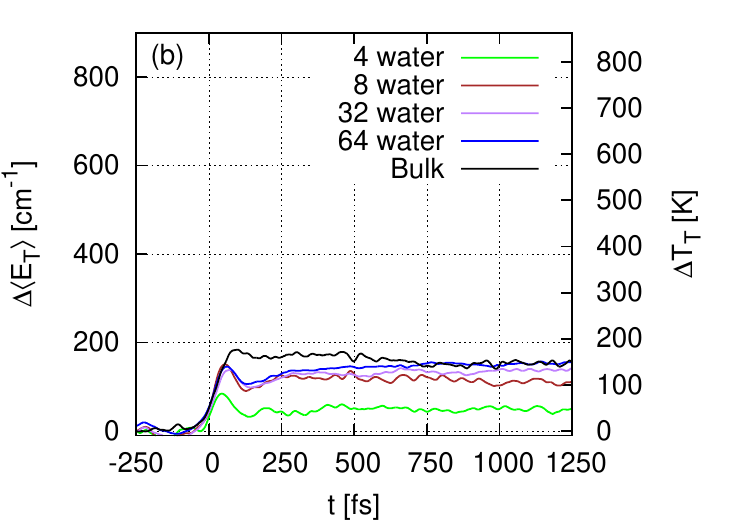}
         }\\%
         \subfigure{%
            \label{fig:cluster-bulk-rot-energy}
            \includegraphics[width=8cm]{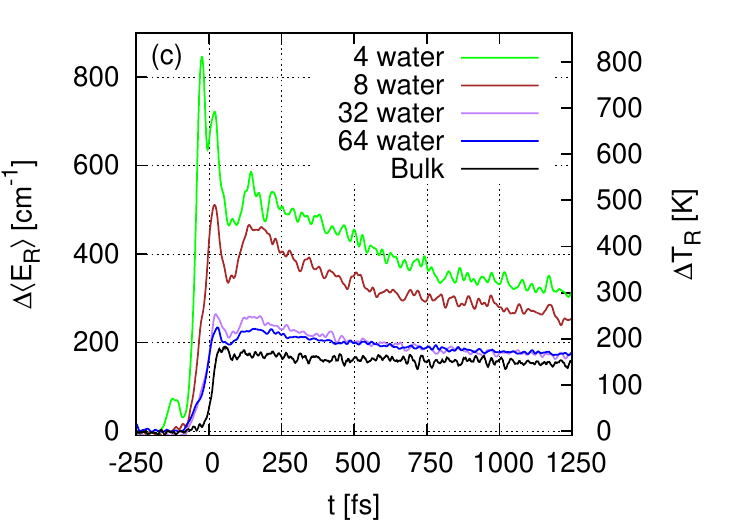}
         }%
     \end{center}
     \caption{    (a) Ensemble averaged total kinetic energy
                      increase ($\Delta\langle E_K\rangle$),
                  (b) translational energy increase
                      ($\Delta\langle E_T\rangle$) and
                  (c) rotational energy increase ($\Delta\langle E_R\rangle$)
                  for the water
                  clusters of sizes 4, 8, 32 and 64 and for bulk water at
                  200~K.
                  Red circles are showing the points t$_{1}$ and t$_{2}$ for 4
                  water cluster only.  t$_{1}$ is the point when significant
                  change happens in $\Delta\langle E_K\rangle$ and t$_{2}$ is the point when
                  $\Delta\langle E_K\rangle$ is maximum.
      }%
    \label{fig:cluster-bulk}
    \end{figure}
For water clusters larger than the dimer the average number of H-bonds
per water molecule increases from $1/2$ to between 3 and 4.
As may already be anticipated,
this has profound consequences for the dynamics of energy increase and energy transfer
in the clusters.
Fig.~\ref{fig:cluster-bulk} illustrates the total kinetic, rotational
and translational energy increase for water clusters of sizes
4, 8, 32, 64 and finally bulk water at the same initial temperature of 200~K.
This temperature was chosen to ensure equilibration of the clusters without
evaporation (loss) of water molecules into their surroundings, which occurs at
higher temperatures.
There are various features of the energy intake dynamics that can now be
understood on the basis of the observations made for the case of the
water dimer.
(i) Smaller clusters feature a larger energy increase immediately after the
THz pulse because they have on average fewer H-bonds per monomer, as seen both
in Figs.~\ref{fig:cluster-bulk-energy} and \ref{fig:cluster-bulk-rot-energy}.
(ii) $\Delta\langle E_R\rangle$ becomes increasingly delayed as the cluster size
increases because the number of surface water molecules with fewer H-bonds
decreases in relation to the cluster's volume. As a consequence,
the time $t_2$, defined as the time of maximum energy gain becomes increasingly
delayed for larger clusters. Exactly the same trend is observed for $t_1$, the
time at which $\Delta\langle E_K\rangle$ starts increasing significantly.
(iii) Equilibration between the $\Delta\langle E_R\rangle$ and
$\Delta\langle E_T\rangle$ components
(cf. Figs.~\ref{fig:cluster-bulk-trans-energy} and
\ref{fig:cluster-bulk-rot-energy}) is achieved for clusters with 32 and 64
water molecules, besides the bulk system. For the smaller clusters with 4 and 8
water molecules, collisional energy transfer stops as the clusters start
expanding, similarly as in the dimer case.
(iv) The total kinetic energy $\Delta\langle E_K\rangle$ decreases after the
pulse for the smaller clusters. Since the total energy of each cluster
is conserved after the pulse, the total kinetic energy
decrease corresponds to an increase in the relative potential energy between
monomers as the small clusters expand. This can be thought of as a sort of
evaporative cooling taking place at the surface of the clusters as outer water
molecules leave the cluster after interaction with the THz pulse.

\subsubsection{Inertial confinement in ultrafast heated clusters}
\begin{figure}[t!]
    \includegraphics[width=8cm]{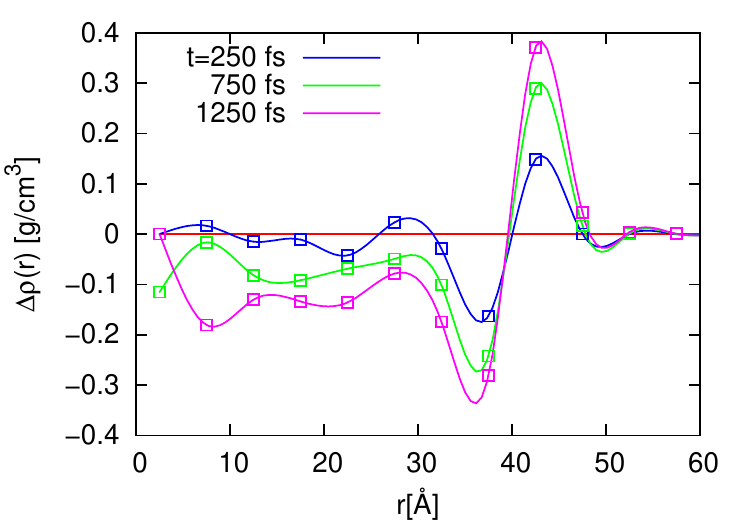}
    \caption{Water density difference ($\Delta \rho$) with respect to the
        density at -250~fs (before the THz pulse) at times 250, 750
        and 1250~fs in the presence of the THz pulse.
        $r$ is the distance to the center of the spherical cluster.
     }%
     \label{fig:density-time}
\end{figure}
In earlier investigations on liquid water interacting with the
THz pulse at a density of
1~g/cm$^3$, constant volume had been assumed~\cite{mishra51,MishraJCPB}. This rests on the
assumption that the number of water molecules in the region interacting with the
THz pulse remains
constant, or in other words, there is no substantial expansion of the system
during and after the interaction with the THz pulse in the time scales of
interest.
As an indication that this may be the case, we already saw in the previous section
that at least the total energy gain converges to the bulk value for clusters
with a few tens
of water molecules.
Here we investigate this assumption in more detail on the basis of
force-field treatment of water, such that several thousand water molecules can
be taken into account.
The dynamics of
an isolated spherical cluster of radius 40~{\AA} and 8843 water molecules, which was
extracted from a thermally equilibrated periodic box of liquid water
at 300~K, were simulated in the presence of the THz pulse. The initial
density in the interior of the cluster was set to 1~g/cm$^3$.

The density as a function of the distance to the center of the cluster was
calculated by binning the radial coordinate in spherical shells of width
5~{\AA} at times $-250$ (before the pulse), $250$, $750$ and $1250$~fs relative
to the center of the THz pulse.
As seen in Fig.~\ref{fig:density-time}, the density varies the most in the outer
5 to 10~{\AA} shell, which diffuses out by also 5 to 10~{\AA}.
The density in the inner regions of the cluster remains quite stable,
decreasing by about 0.15~g/cm$^3$ during the first picosecond after the pulse.
Larger clusters may still lead to a more effective inertial confinements, and it appears
justified to assume a constant density in simulations of heated bulk water at the time
scales of interest.

\subsection{THz pulse interaction with bulk water}
\label{ssec:bulk}
\begin{figure}[h!]
    \includegraphics[width=8cm]{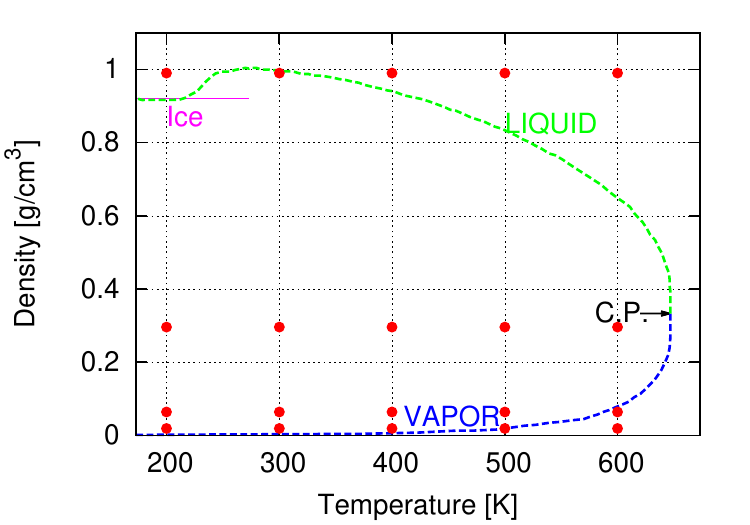}
    \caption{Temperature (T)-density ($\rho$) phase diagram of water.  C.P. is
        the critical point. Red dots are the states of water in the
        phase diagram that have been used to explore the effect of THz pulse.
     }%
     \label{fig:phase-diagram}
\end{figure}
%

\begin{figure*}[h!]
      \begin{center}
      \subfigure{%
            \label{fig:temperature-effect-rot}
            \includegraphics[width=0.4\textwidth]{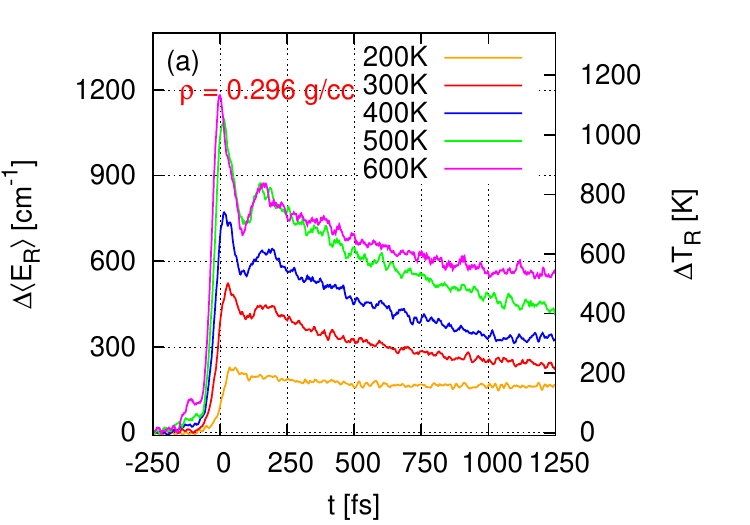}
         }%
       \subfigure{%
            \label{fig:temperature-effect-rot-019}
            \includegraphics[width=0.4\textwidth]{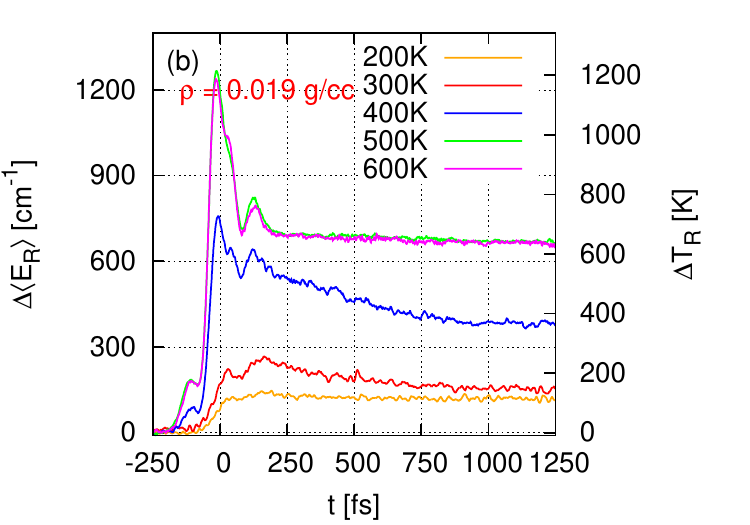}
         }\\%
       \subfigure{%
            \label{fig:temperature-effect-trans}
            \includegraphics[width=0.4\textwidth]{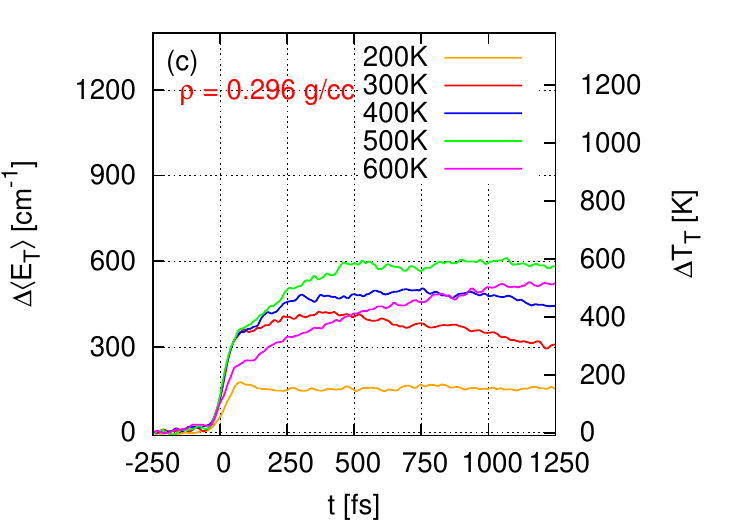}
       }%
       \subfigure{%
            \label{fig:temperature-effect-trans-019}
            \includegraphics[width=0.4\textwidth]{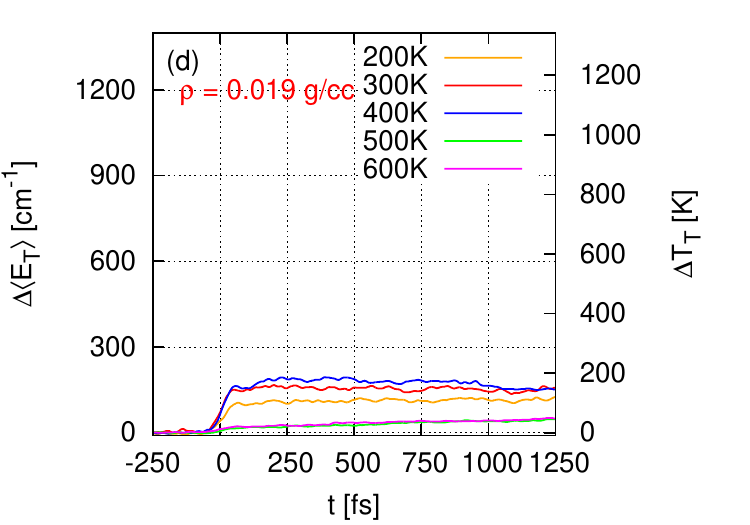}
       }\\%
       \subfigure{%
            \label{fig:temperature-effect-tot}
            \includegraphics[width=0.4\textwidth]{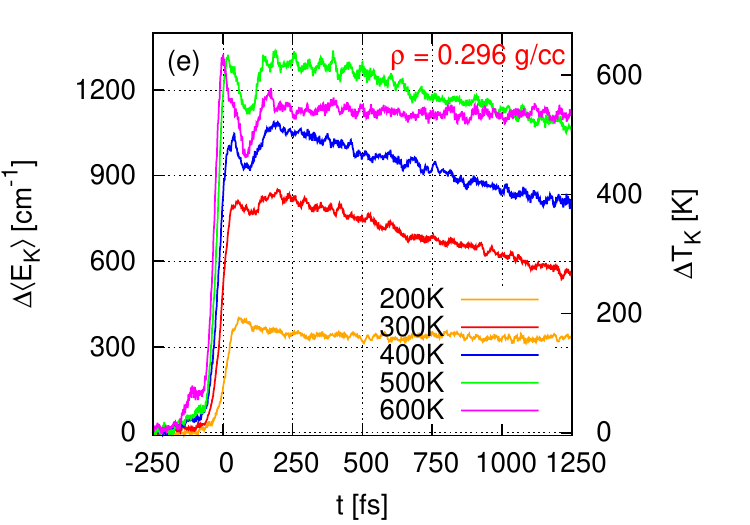}
         }%
       \subfigure{%
            \label{fig:temperature-effect-tot-019}
            \includegraphics[width=0.4\textwidth]{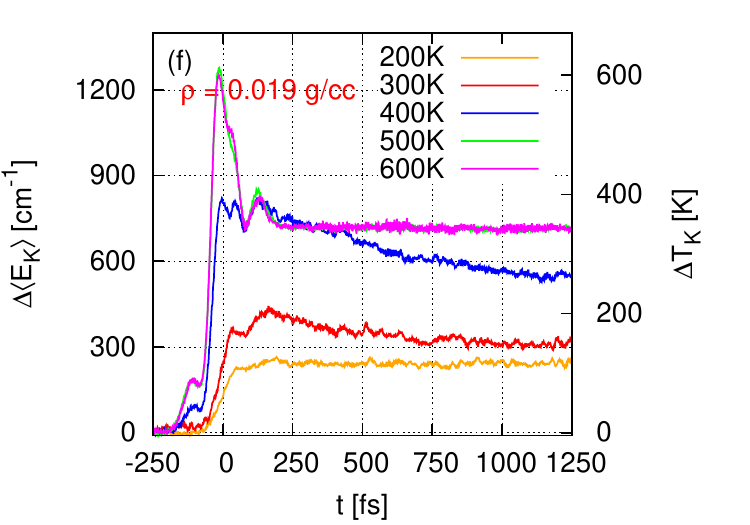}
         }%
     \end{center}
     \caption{Total kinetic energy ($\Delta \langle E_K\rangle$), translational
         energy ($\Delta \langle E_T\rangle$) and rotational energy ($\Delta
         \langle E_R\rangle$) per water molecule at different temperatures for
         densities 0.296 g/cm$^3$ (a,c,e) and 0.019 g/cm$^3$ (b,d,f).
      }%
    \label{fig:temperature-effect-total}
\end{figure*}
\begin{figure*}[h!]
    \begin{minipage}{\textwidth}
      \begin{center}
         \subfigure{%
            \label{fig:snap-200-0}
            \includegraphics[width=0.46\textwidth]{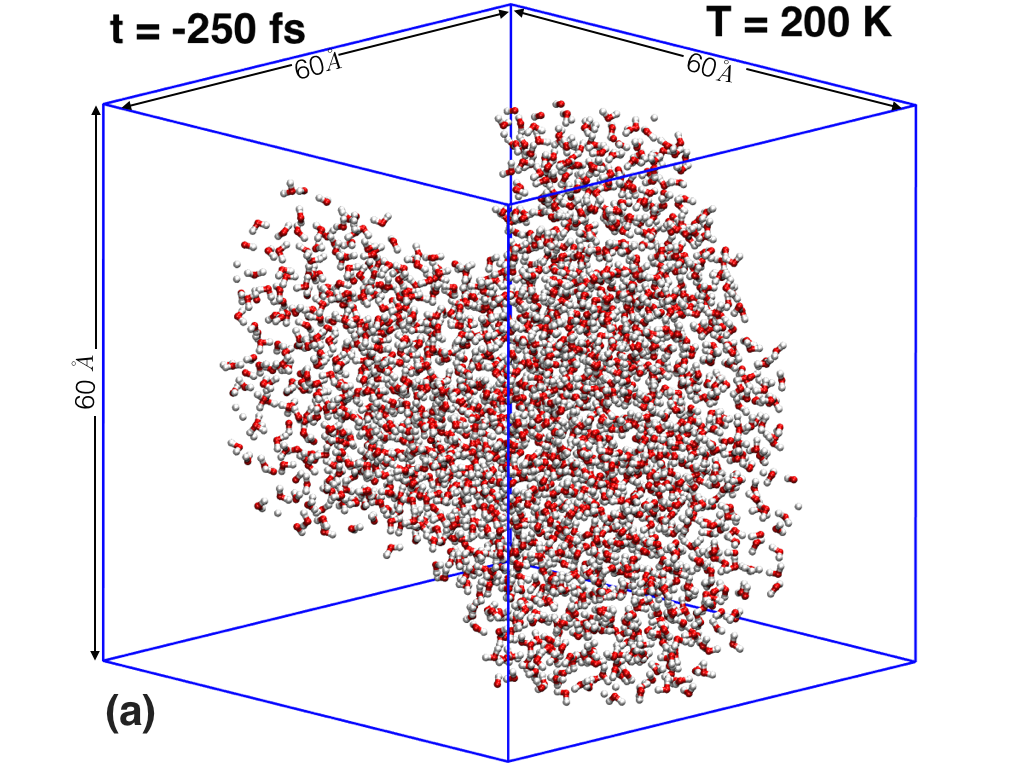}
         }%
         \subfigure{%
            \label{fig:snap-200-1500}
            \includegraphics[width=0.4\textwidth]{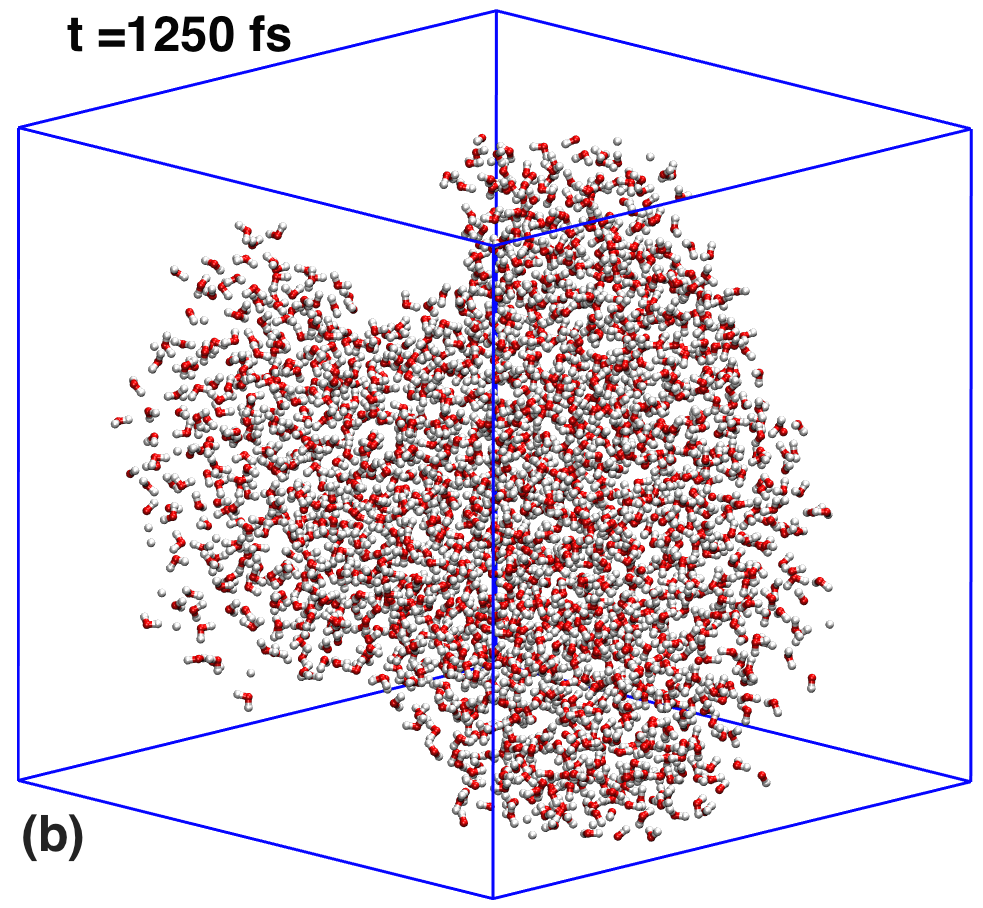}
         }\\%
         \subfigure{%
            \label{fig:snap-400-0}
            \includegraphics[width=0.4\textwidth]{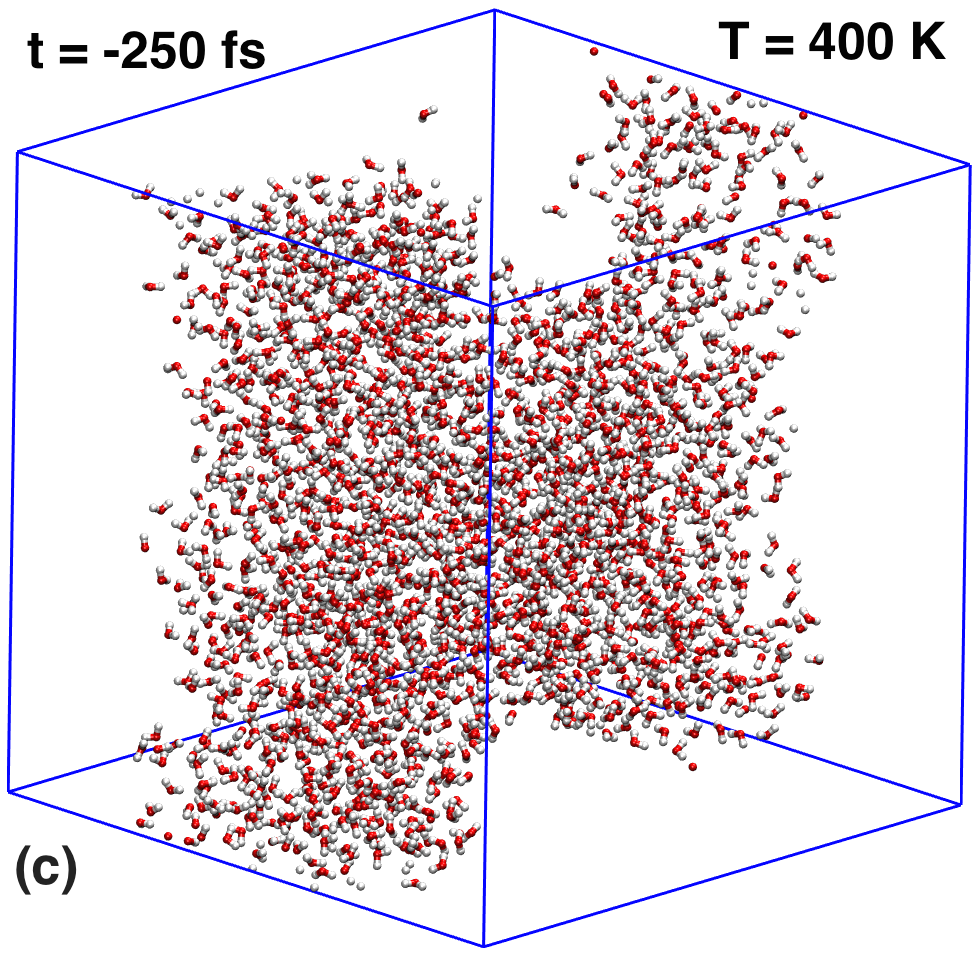}
         }%
         \subfigure{%
            \label{fig:snap-400-1500}
            \includegraphics[width=0.4\textwidth]{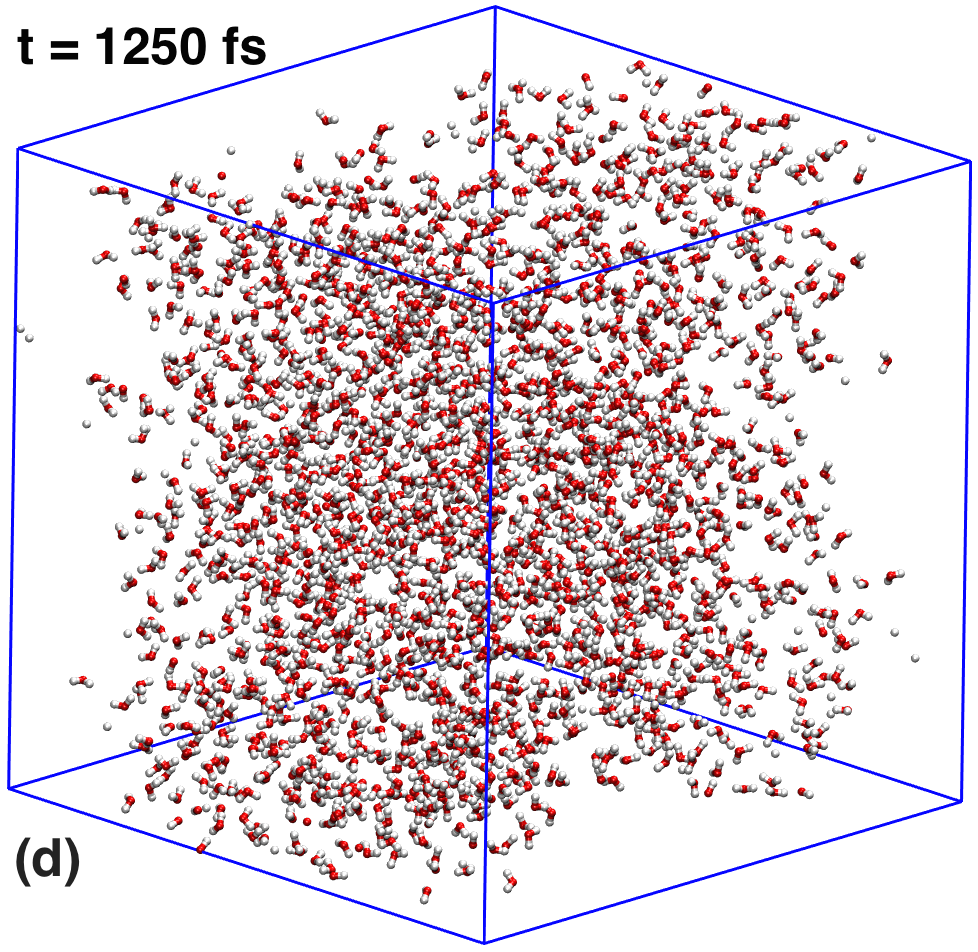}
         }\\%
         \subfigure{%
            \ \label{fig:snap-600-0}
            \includegraphics[width=0.4\textwidth]{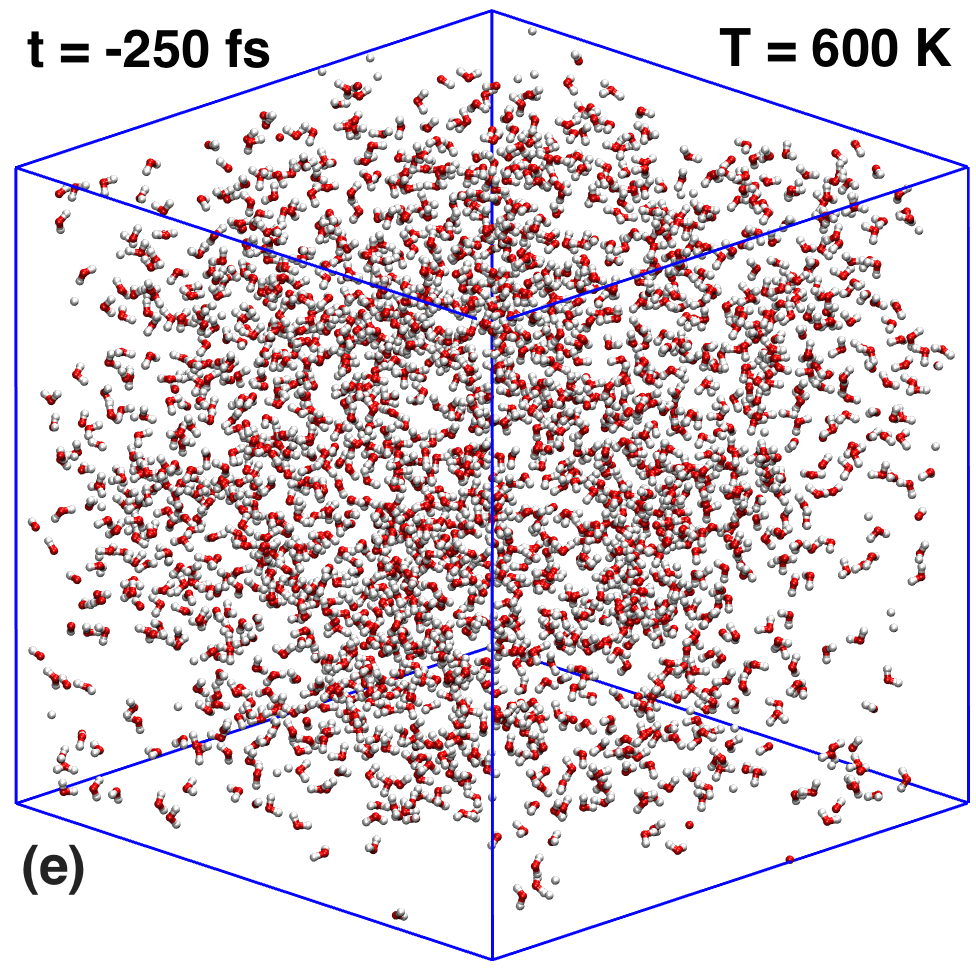}
         }%
         \subfigure{%
            \label{fig:snap-600-1500}
            \includegraphics[width=0.4\textwidth]{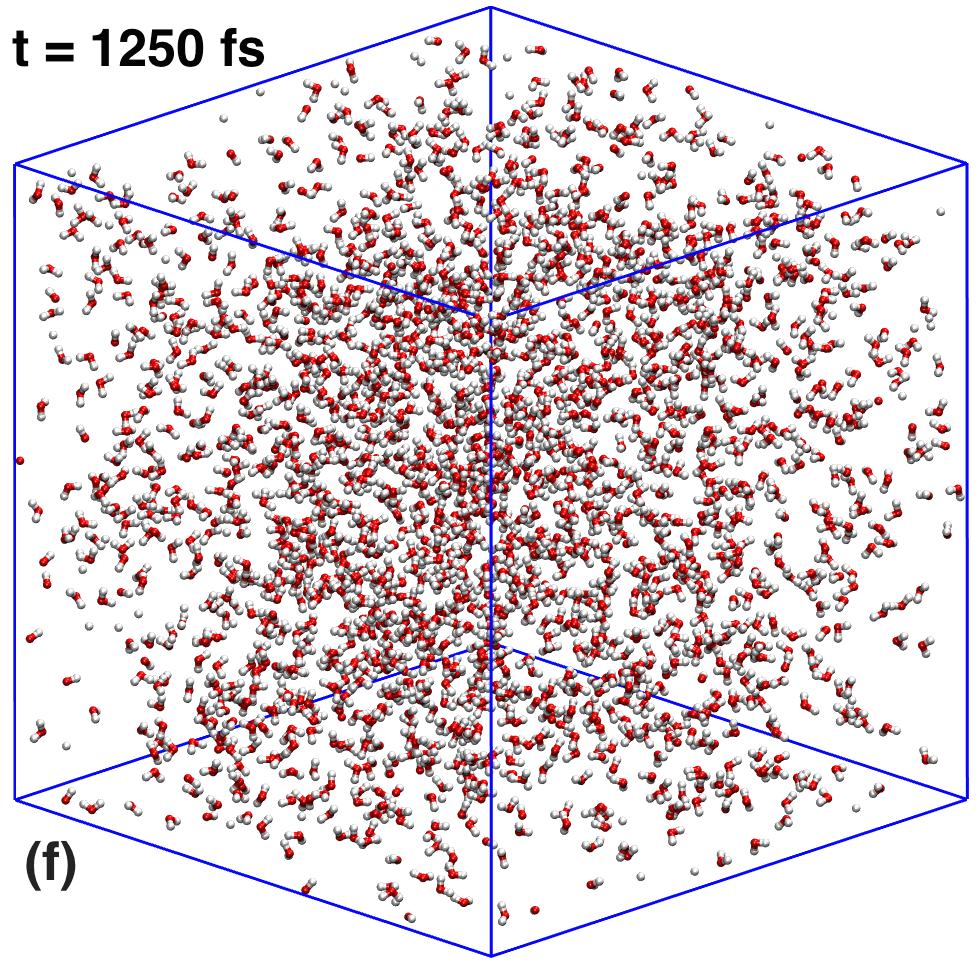}
         }\\%
     \end{center}
     \caption{Snapshots of simulation boxes with density 0.3 g/cm$^3$
         and different
         temperatures (T= 200~K, 400~K and 600~K) before the pulse, $t=-250$~fs
         (a,c,e) and after the pulse, $t=1250$~fs (b,d,f).
      }%
    \label{fig:snapshot-temperature-effect}
    \end{minipage}
    \end{figure*}

We have discussed above how an intense, sub-cycle THz pulse interacts with
an isolated water molecule and with water clusters of various sizes.
Summarizing, an isolated water molecule interacts with the pulse through its
permanent dipole moment thus gaining a large amount of rotational energy.
Only when several water molecules are in close proximity can part of the
rotational energy be transformed into relative translational energy owing to the
coupling between rotational and relative translational motion of the water
monomers in the cluster.

In the following, we explore the interaction of the THz pulse and
the subsequent energy transfer dynamics in bulk water at different constant
densities and equilibrated at different initial temperatures.
Regarding initial temperature, this parameter is easily controllable in
experimental realizations.
As for density, we have already shown that, on the time scales of interest,
the inner part of the cluster remains at its
initial density during the interaction with the pulse and is inertially confined
by the outer shell. Therefore, simulations at the constant density (i.e. constant
volume) of the liquid phase, 1~g/cm$^3$, are justified.
Moreover, the macroscopic density under real physical confinement of the
interaction volume by e.g. hard walls can be maintained during and after the heating-up process.
Therefore, we consider various densities ranging from very low up
to normal liquid density.
Depending on the density and initial temperature of the confined system in equilibrium,
the aggregation state of water may vary from gas phase to liquid or amorphous
solid phases and the coexistence of the two.
This is illustrated in the saturation diagram of pure water shown in
Fig.~(\ref{fig:phase-diagram})~\cite{wag02:387}.
In the regions inside the curve vapor and liquid water coexist. Outside
the curve only one phase is present, either water vapor or condensed water.
The dynamics of energy transfer from the THz pulse and the subsequent energy
redistribution among rotational and translational degrees of freedom are
strongly dependent on the possible aggregation states found in the diagram.
Those initial conditions at which simulations have been performed in this work are marked
with a red dot in Fig.~(\ref{fig:phase-diagram}).

%
%


There are three distinctive features of the
heating-up dynamics of water clusters that vary strongly as a
function of the size of the cluster.
{\em First},
for isolated water molecules and for clusters up to 8 monomers the amplitude
profile of the THz pulse is imprinted in the rotational energy of the water
molecules. This is indicative of none or few hydrogen bonds per monomer as
compared to the bulk or to larger clusters. As a consequence, water molecules
are rotationally accelerated and shortly thereafter slowed down in the presence
of the THz field before they can have significant collisions with neighboring
monomers. This can be observed in Figs.~\ref{fig:water-monomer-energy} and
\ref{fig:cluster-bulk}c.
{\em Second}, an equilibriation between the rotational and translational
components of the kinetic energy is only achieved for clusters after a certain
minimum size. In our simulations we find equilibriation between rotational and
translational energy components
for clusters of 32 or more water molecules. Smaller clusters fragment before
equilibriation through collisions can complete. An extreme example of this
corresponds to
the water dimer and tetramer.
{\em Third}, the total kinetic energy of the clusters with four and eight
monomers decreases during the first picosecond as the clusters expand, which is
mirrored by the corresponding increase of potential energy as the monomers
separate from each other against attractive electrostatic interactions.
This effect is not observed for larger
aggregates, which mostly maintain their structure.

\begin{figure*}[h!]
    \begin{minipage}{\textwidth}
      \begin{center}
         \subfigure{%
            \label{fig:snap-60-0}
            \includegraphics[width=0.4\textwidth]{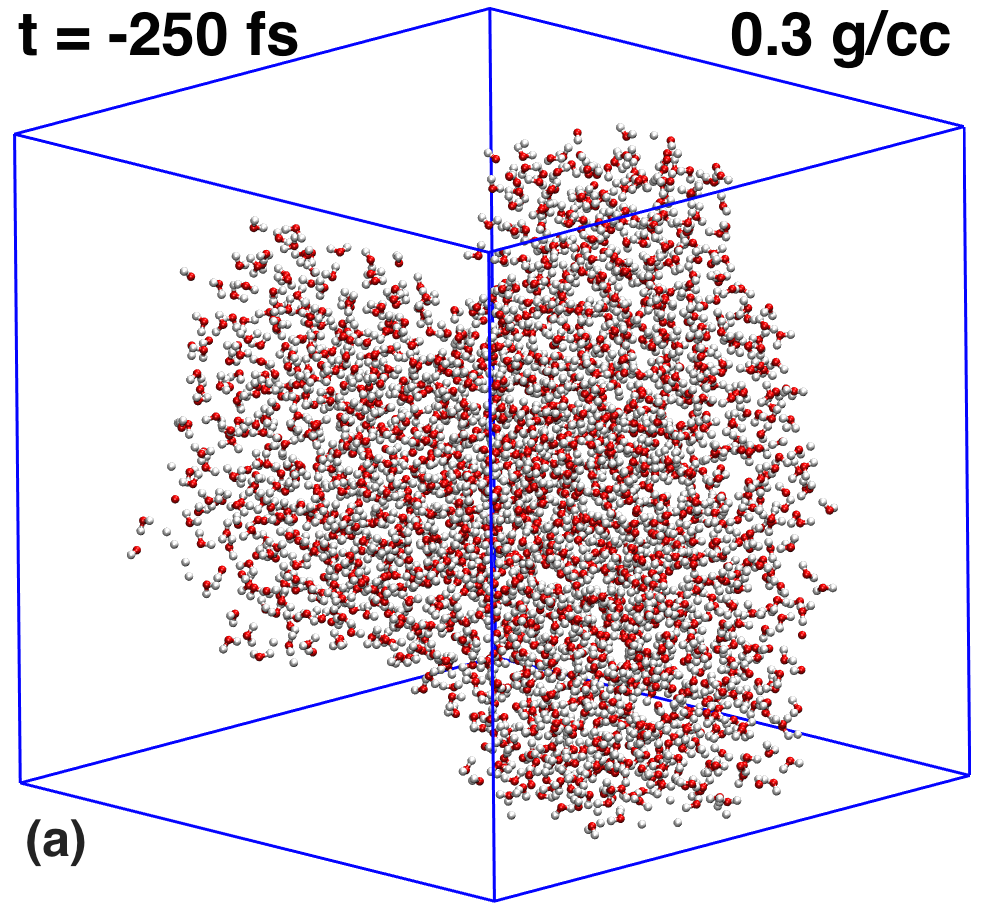}
         }%
         \subfigure{%
            \label{fig:snap-60-1500}
            \includegraphics[width=0.4\textwidth]{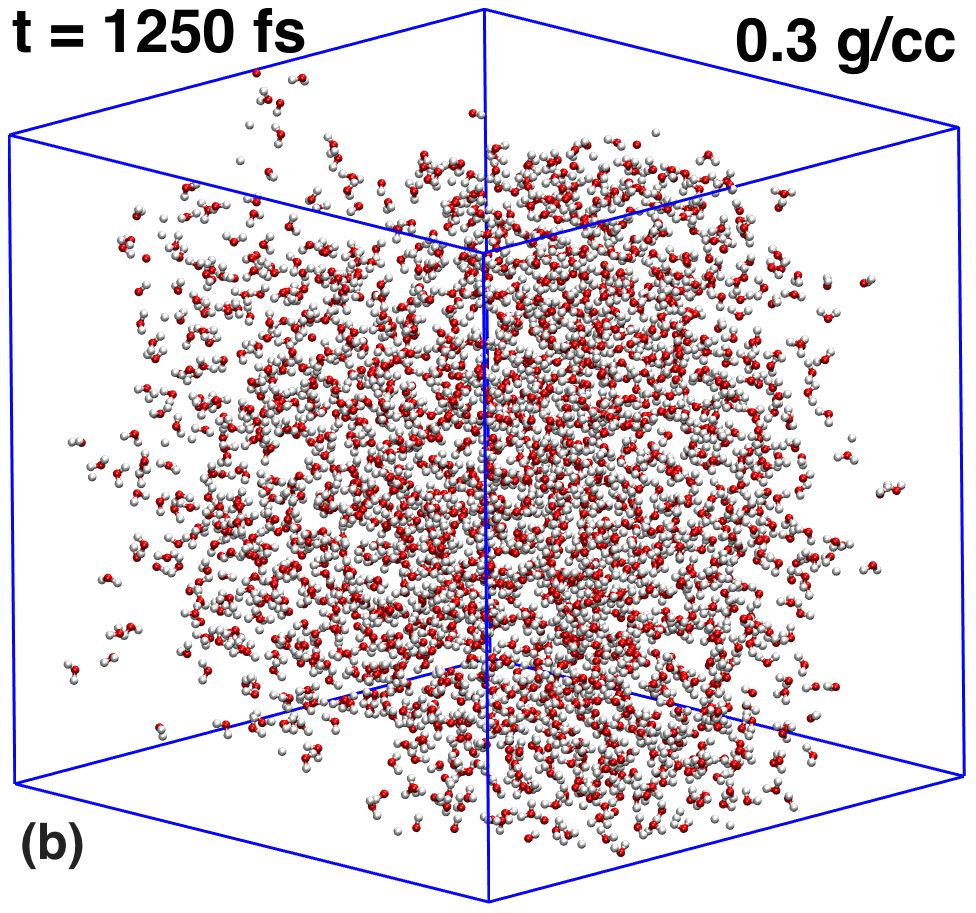}
         }\\%
         \subfigure{%
            \ \label{fig:snap-100-0}
            \includegraphics[width=0.4\textwidth]{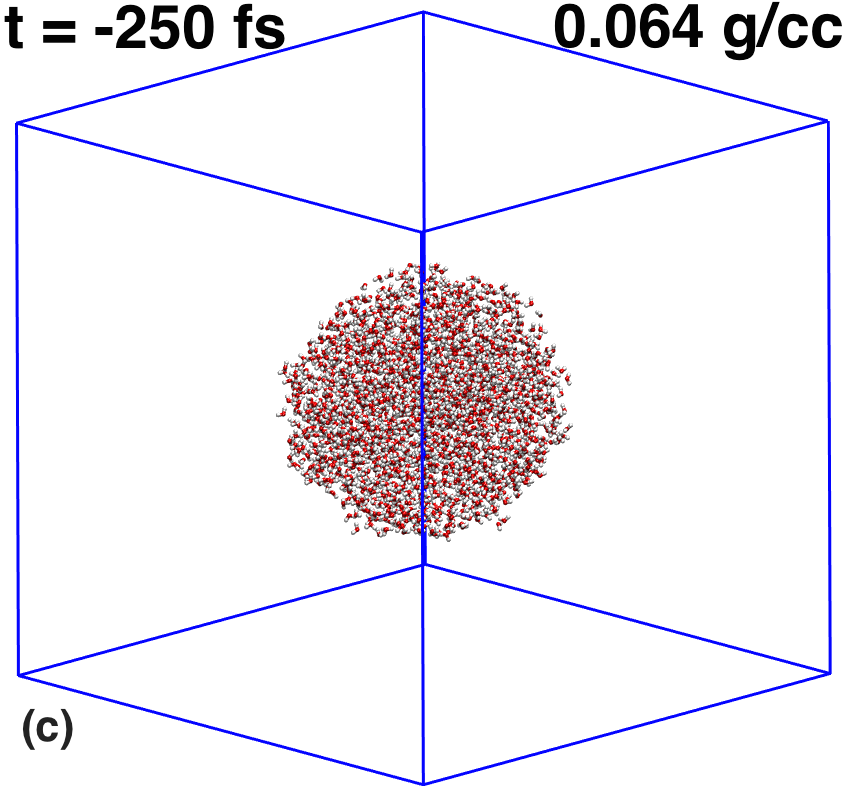}
         }%
         \subfigure{%
            \label{fig:snap-100-1500}
            \includegraphics[width=0.4\textwidth]{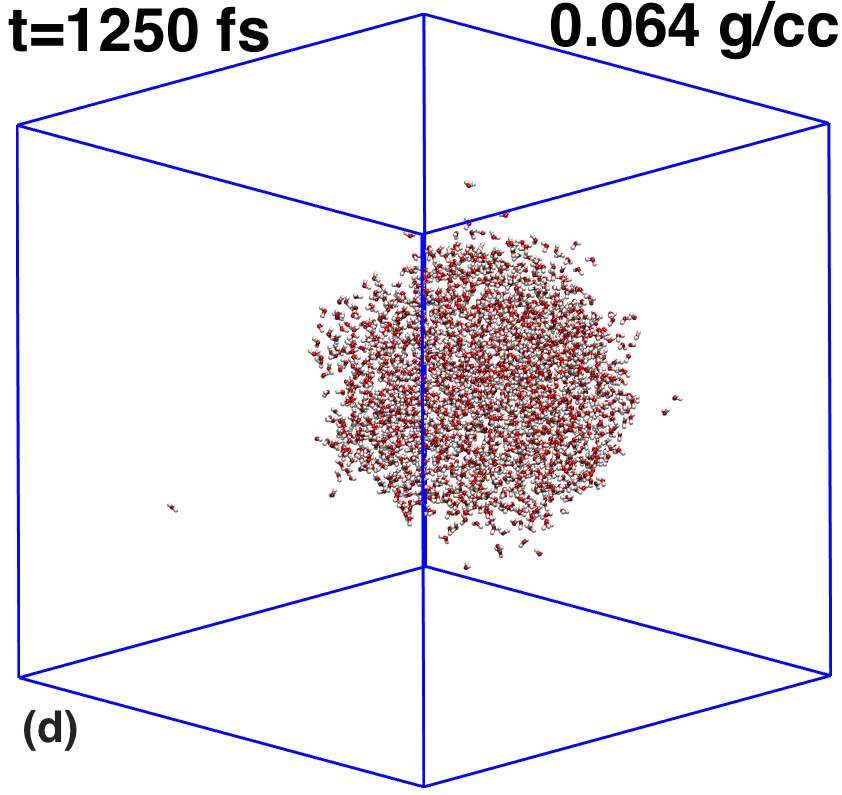}
         }\\%
         \subfigure{%
            \ \label{fig:snap-150-0}
            \includegraphics[width=0.4\textwidth]{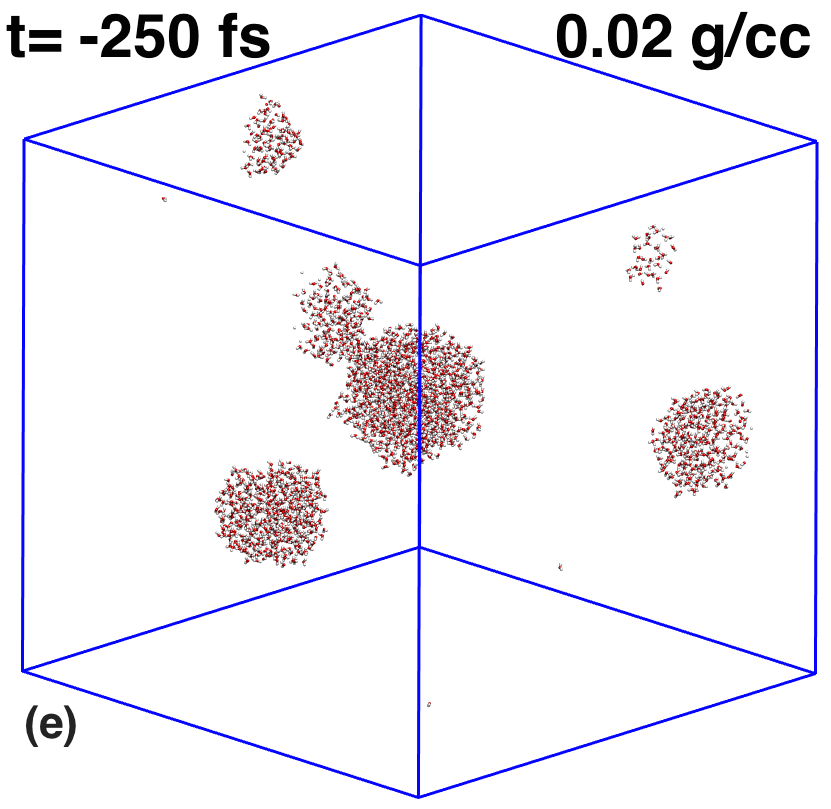}
         }%
         \subfigure{%
            \label{fig:snap-150-1500}
            \includegraphics[width=0.4\textwidth]{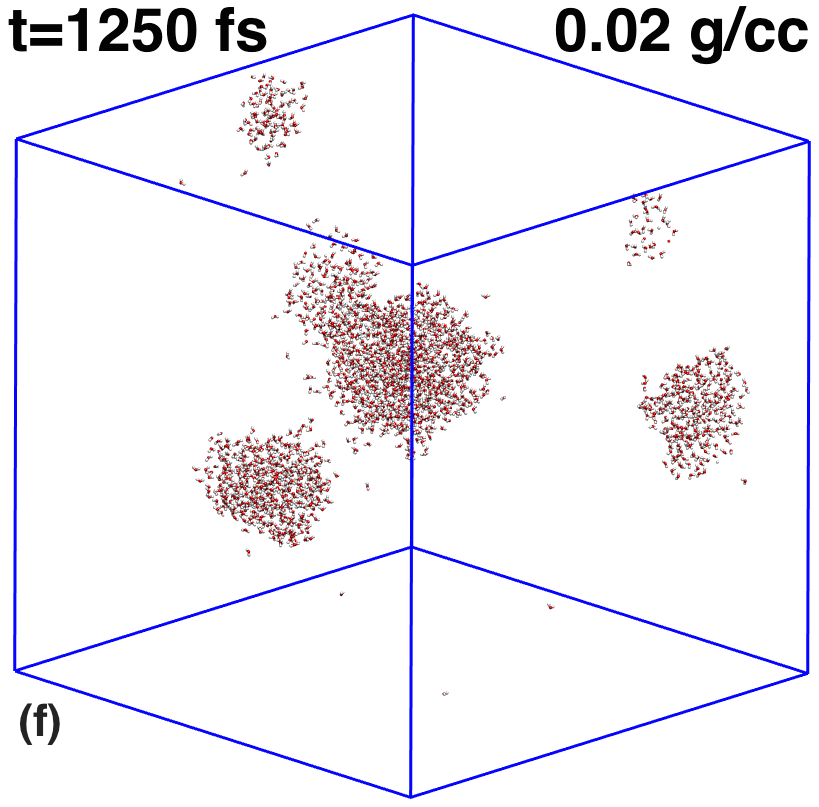}
         }\\%
     \end{center}
     \caption{Snapshots of water systems at T=300~K and different densities
              (0.3 g/cm$^3$, 0.064 and 0.02 ~g/cm$^3$)
              before the pulse, $t=-250$~fs (a,c,e) and after the pulse,
              $t=1250$~fs (b,d,f).
      }%
    \label{fig:snapshot-density-effect}
    \end{minipage}
    \end{figure*}

These basic features can be recognized when letting
bulk water interact with the THz pulse at different densities and initial
temperatures.
Snapshots of 2048 water molecules at a density of 0.296~g/cm$^3$ for different
initial temperatures of 200, 400 and 600~K are shown in
Fig.~\ref{fig:snapshot-temperature-effect}. The system at 200~K
corresponds essentially to a piece of amorphous ice with small internal energy.
The system at 400~K is made of a large droplet of hot liquid water with
smaller clusters and monomers surrounding it in the vapor state,
whereas the system at 600~K is made of almost
homogeneously distributed water molecules with a large mobility and collisional
rate.
This leads immediately to very different heating-up dynamics. The colder system
gains in total less energy due to the more hindered water monomers. The
equilibration between rotational and translational energy is however almost
instantaneous due to the tight interactions between the monomers.
As the temperature increases the amplitude profile of the THz pulse manifests
more strongly in the rotational energy component because the interactions
between neighboring water molecules become progressively weaker, as seen in
Fig.~\ref{fig:temperature-effect-total}a.
Another consequence of larger temperatures is that
the energy transfer between the rotational and translational
components of the kinetic energy is not instantaneous anymore as compared
to colder temperatures (cf.
Figs.~\ref{fig:temperature-effect-total}c and
\ref{fig:temperature-effect-total}d).  The slowest energy transfer to
translational motion occurs at 600~K as a consequence of the larger separation
and weaker interactions between water molecules.
A decrease of the total kinetic energy shortly after the pump pulse occurs for
systems initially between 300 and 500~K due to substantial evaporation of
monomers from the outer shells of the liquid portion. This can be seen by
comparing the snapshots in
Figs.~\ref{fig:snapshot-density-effect}a,
\ref{fig:snapshot-density-effect}c and
\ref{fig:snapshot-density-effect}e
with those in Figs.~\ref{fig:snapshot-density-effect}b,
\ref{fig:snapshot-density-effect}d and
\ref{fig:snapshot-density-effect}f, respectively.

As compared to the previous examples, a reduced density of 0.019~g/cm$^3$ leads
to differences mostly for the high temperature cases. At T$=500$~K and T=$600$~K
the water molecules are homogeneously distributed across the available volume
and at a large average distance from one another. This results in dynamics
of the rotational degrees of freedom similar to those of isolated monomers and
characterized by a
slow rate of energy transfer to translational degrees of freedom due to
infrequent collisions. This can be seen by comparing the increase of
translational energy shown in
Figs.~\ref{fig:temperature-effect-total}c and
\ref{fig:temperature-effect-total}d.
At this density, only the colder systems at T$=200$~K and T$=300$~K achieve
energy equipartition between rotational and translational degrees of freedom because
most of the water molecules are condensed forming ice and liquid droplets, as seen in
Figs.~\ref{fig:snapshot-density-effect}c and
\ref{fig:snapshot-density-effect}e.
This leads to coupled hindered rotational and translational
motion in the condensed phase and the corresponding fast energy transfer.

    \begin{figure*}[h!]
      \begin{center}
         \subfigure{%
            \label{fig:density-energy-rot-300}
            \includegraphics[width=0.4\textwidth]{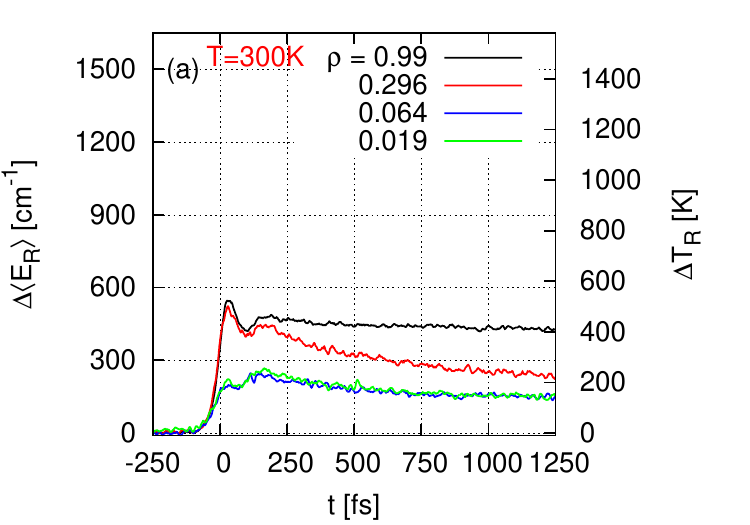}
         }%
         \subfigure{%
            \label{fig:density-energy-rot}
            \includegraphics[width=0.4\textwidth]{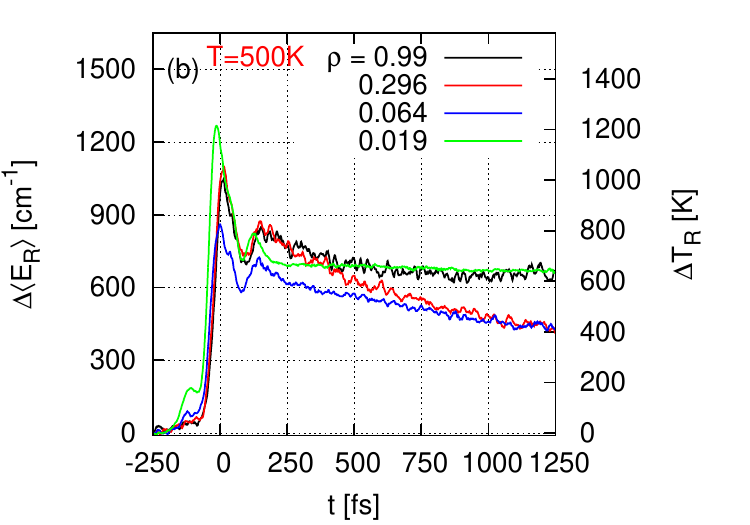}
         }\\%
        \subfigure{%
            \label{fig:density-energy-trans-300}
            \includegraphics[width=0.4\textwidth]{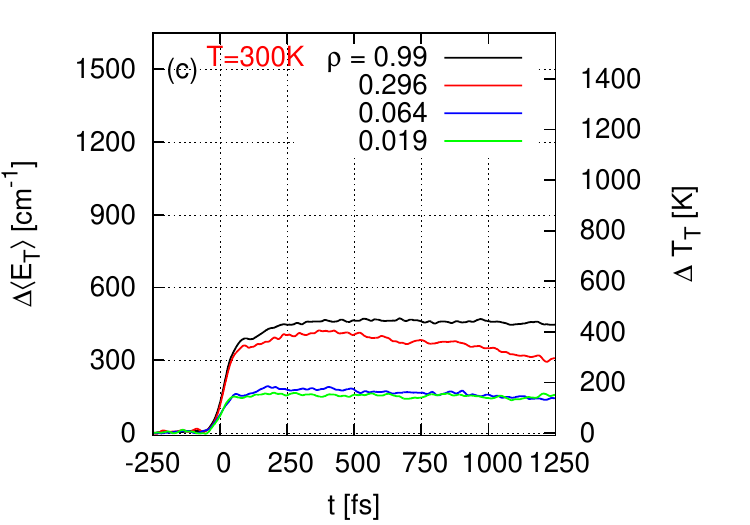}
         }%
         \subfigure{%
            \label{fig:density-energy-trans}
            \includegraphics[width=0.4\textwidth]{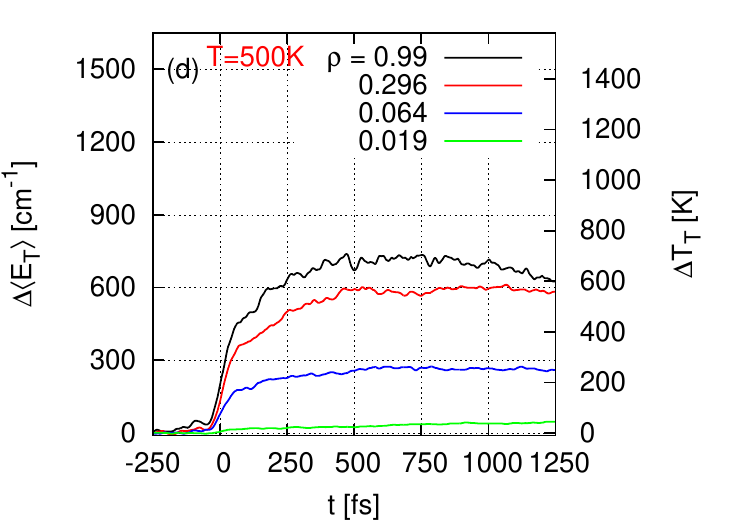}
         }\\%
         \subfigure{%
            \label{fig:density-energy-tot-300}
            \includegraphics[width=0.4\textwidth]{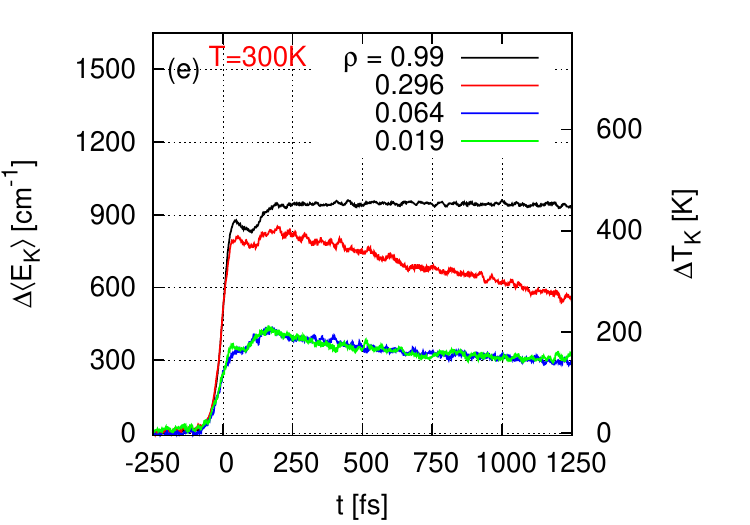}
         }%
         \subfigure{%
            \label{fig:density-energy-tot}
            \includegraphics[width=0.4\textwidth]{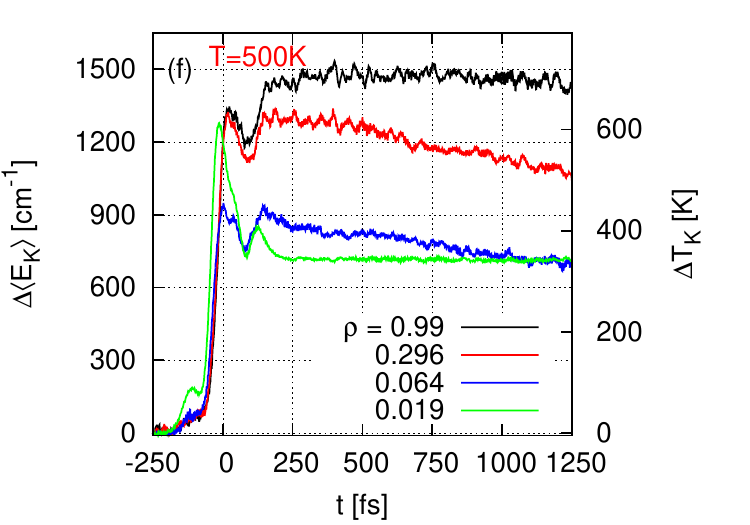}
         }%
    \end{center}
     \caption{ Total kinetic energy ($\Delta \langle E_K\rangle$), translational
         energy ($\Delta \langle E_T\rangle$) and rotational energy ($\Delta
         \langle E_R\rangle$) per water molecule at different densities
         (in g/cm$^3$) for
         initial temperatures 300~K (a,c,e) and 500~K (b,d,f) of the system.
      }%
    \label{fig:density-effect}
    \end{figure*}

At temperatures T=500~K and T=600~K the amount and rate of energy transfer
from the THz pulse to
rotational motion is relatively independent of the density of the system, as
can be seen in Fig.~\ref{fig:density-effect}b. Independently of the density, the
high temperatures result in weak hydrogen bonding and the rotational degrees of
readily respond to the THz pulse.
However, the collisional energy transfer from the rotational to the
translational degrees of freedom at these high temperatures is strongly
dependent on the density of the system, as illustrated in
Fig.~\ref{fig:density-effect}d.
At low densities almost no collisions occur whereas at 1~g/cm$^3$ the rotational
and translational temperatures are equilibriated within 1~ps.

As a final remark, the simulations of the interaction of the THz pulse with liquid
water at an initial temperature of 300~K and density of 1~g/cm$^3$ can be
compared with our previous {\em ab initio} molecular dynamics
results~\cite{mishra51,MishraJCPB}. The total kinetic energy increase
$\Delta\langle E_K\rangle$ per water monomer is found to be about a factor 2
larger in the {\em ab initio} case than with the present rigid
force-field approach. The main difference is
related to the direct energy intake of the internal vibrational degrees of
freedom of the water monomers during the THz pulse and which is related to
polarization effects not captured by rigid force-fields. The mechanism of energy
transfer to rotational and translational degrees of freedom remains however
the same and the amount of energy transferred to these modes is in both types of
simulations qualitatively agrees to within 20\%.

\section{Conclusion}
\label{sec:concl}

We have discussed in detail the mechanisms by which
a highly intense (5$\times$10$^{12}$~W/cm$^{2}$) and ultrashort (fwhm 141~fs)
THz pulse couples to clusters of water molecules and to bulk water and
transfers a large amount of energy to the system.
The fundamental coupling mechanism between the THz pulse and the water molecules
is through the permanent dipole moment of those with the electric field
component of the pulse.
Therefore, the THz pulse transfers energy only to rotational motion of isolated
water molecules and
translational motion of the center of mass remains unaffected
due to electrical neutrality of the molecules.
For clusters with two or more molecules, coupling between hindered rotations
and translations exists and the translational motions also gain energy during and
after interaction with the THz pulse.
Inspection of the potential energy curve of the $\angle$OOH angle of the water
dimer in the presence
of a static electric field shows
small changes for a electric field amplitude up to 0.087~V/{\AA}. Beyond this
field strength, a small variation in the electric field will induce a significantly large
change in the potential energy curve.
At a electric field amplitude of 0.614 V/{\AA}, the potential energy curve shows that the electric
field can break the
hydrogen bond in the water dimer and bring the permanent dipole moments of the
two monomers to orient with the polarization axis of the electric field.

We studied the interaction of confined water at well defined initial temperature
and different densities. Even relatively small clusters lead to inertial
confinement of the water molecules below the most external shells, which
justifies investigations of bulk water at constant density of
1~g/cm$^3$. We also investigated lower densities, which can result from physical
confinement either by macroscopic physical walls or even in nanostructures.
High temperatures of 500 and 600~K lead to rotational dynamics similar to
isolated water molecules. Lower temperatures, in which the interactions between
water monomers are stronger, lead to a smaller energy transfer from the THz
pulse.
The density of the system plays a key role in the collisional energy transfer
between rotational and translational motion.
For large temperatures, a
large density leads to a high collision rate and quick equilibriation whereas a
small density prevents collisions and no equilibriation can occur.
For small temperatures, a large fraction of the system is in a condensed state
and equilibriation between rotational and translational motion occurs almost
instantaneously.

These results constitute a relevant step towards studies of
thermally activated chemical
and biological processes, in which the water medium can be used as a channel to
deliver the energy of a THz pump pulse to the molecules of interest. The trends
identified in terms of temperature and density of the system
open the door to the
design of strategies in which the heating-up rate of substances dissolved in water
can be controlled by such thermodynamic parameters.

\begin{acknowledgments}
   We are grateful to the Virtual Institute of the Helmholtz Association
   "Dynamic Pathways in Multidimensional Landscapes" for financial support.
\end{acknowledgments}

\bibliography{references}



\end{document}